\documentclass[12pt,english]{article}
\usepackage{mathptmx}
\usepackage{courier}

\usepackage[T1]{fontenc}
\usepackage[latin9]{inputenc}
\usepackage[a4paper]{geometry}
\usepackage{color}
\usepackage{babel}
\usepackage{array}
\usepackage{float}
\usepackage{amsmath}
\usepackage{amssymb}
\usepackage{graphicx}
\usepackage{setspace}
\usepackage[unicode=true,pdfusetitle,
 bookmarks=true,bookmarksnumbered=false,bookmarksopen=false,
 breaklinks=true,pdfborder={0 0 0},backref=false,colorlinks=false]
 {hyperref}
\makeatletter

\providecommand{\tabularnewline}{\\}
\floatstyle{ruled}
\newfloat{algorithm}{tbp}{loa}
\providecommand{\algorithmname}{Algorithm}
\floatname{algorithm}{\protect\algorithmname}

\usepackage{url}

\vfuzz \hfuzz
\definecolor{darkblue}{rgb}{1,0,0}
\definecolor{darkred}{rgb}{0.8,0,0}
\definecolor{black}{rgb}{0,0,0}

\RequirePackage{url}
\hypersetup{
breaklinks=true,
bookmarksopen=true,
bookmarksnumbered=true,
linkcolor=black,
urlcolor=black,
citecolor=darkblue,
colorlinks=false}

\@ifundefined{showcaptionsetup}{}{%
 \PassOptionsToPackage{caption=false}{subfig}}
\usepackage{subfig}
\makeatother
\date{}
\begin{document}

\title{Proactive QoE Provisioning in Heterogeneous Access Networks using
Hidden Markov Models and Reinforcement Learning}

\author{Karan Mitra%
\thanks{K. Mitra (Corresponding author) is with Lule\aa{} University of Technology (LTU), Sweden. C.
Åhlund is with LTU, Sweden. A. Zaslavsky is with CSIRO, Australia.
Saguna is with LTU, Sweden.%
}, Christer Åhlund, Arkady Zaslavsky and Saguna}
\maketitle
\begin{abstract}
\textbf{Quality of Experience (QoE) provisioning in heterogeneous
access networks (HANs) can be achieved via handoffs. The current approaches
for QoE-aware handoffs either lack the availability of a network path
probing method or lack the availability of efficient methods for QoE
prediction. Further, the current approaches do not explore the benefits
of proactive QoE-aware handoffs such that user's QoE is maximized
by learning from past network conditions and by actions taken by the
mobile device regarding handoffs. In this paper, our contributions
are two-fold. First, we propose, develop and validate a novel method
for QoE prediction based on passive probing. Our method is based on
hidden Markov models and Multi-homed Mobility Management Protocol
which eliminates the need for additional probe packets for QoE prediction.
It achieves the average QoE prediction accuracy of 97\%. Second, we
propose, develop and validate a novel reinforcement learning based method for proactive
QoE-aware handoffs. We show that our method outperforms existing approaches
by reducing the number of vertical handoffs by 60.65\% while maintaining
high QoE levels and by extending crucial functionality such as passive
probing mechanisms.}
\end{abstract}

\begin{table}[H]
\caption{List of notations used in this paper.}

\centering{}%
\begin{tabular}{|c|c|}
\hline 
Notation & Meaning\tabularnewline
\hline 
\hline 
$i$ & a network interface\tabularnewline
\hline 
$\Gamma(\bullet)$ & handoff function\tabularnewline
\hline 
$E (\bullet)$ & expected reward function\tabularnewline
\hline 
$\Delta$ & average number of handoffs\tabularnewline
\hline 
$\Phi$ & QoE prediction accuracy\tabularnewline
\hline 
$t$ & time-stamp/decision epoch\tabularnewline
\hline 
$A$ & a set of actions\tabularnewline
\hline 
$a^{t}$ & context attribute at $t$\tabularnewline
\hline 
$w_{j}$ & weight of an attribute\tabularnewline
\hline 
$E^{t}$ & an observation/evidence at $t$\tabularnewline
\hline 
$\mu$ & mean\tabularnewline
\hline 
$\sigma^{2}$ & variance\tabularnewline
\hline 
$TM$ & transition matrix\tabularnewline
\hline 
$p$$ $ & initial state distribution/prior\tabularnewline
\hline 
$\Theta$ & model parameters\tabularnewline
\hline 
$\Re(\bullet)$ & a reward function\tabularnewline
\hline 
$\pi$ & a policy\tabularnewline
\hline 
$S$ & set of system states\tabularnewline
\hline 
$\gamma$ & discount factor\tabularnewline
\hline 
$\alpha$ & learning rate\tabularnewline
\hline 
\end{tabular}
\end{table}

\section{Introduction}

Mobile devices such as tablets and smartphones can connect to heterogeneous
access networks (HANs) using a plethora of wireless technologies such
as WiFi and 4G. Users using their mobile devices may roam in HANs
while moving from one wireless network to another. Users usually associate
some expectations while accessing applications on their mobile devices
\cite{sung}. These expectations along with their cognitive and behavioural
states and network quality of service (QoS), may dictate their quality
of experience (QoE) \cite{sung,1631338,mitratmc2014}. If users are
satisfied with their QoE, they may stay with the current telecommunication
operator else they may switch to a new operator. For instance, in
2011, Vodafone Australia nearly lost 440,000 customers to different
operators such as Telstra and Optus due to customer dissatisfaction%
\footnote{http://www.itnews.com.au/News/290168,vodafone-australia-churn-nears-half-a-million-for-2011.aspx (retrieved 03/09/15).
}. Telecommunication operators are interested in maximizing their revenue
by trying to retain their customers. On the other hand, users consider
operators that meet their QoE. Thus, there is a need to facilitate
QoE while users roam in HANs such that both telecommunication operators
and users are satisfied \cite{koumaras2012telsys}. 

QoE provisioning in HANs can be achieved through handoffs \cite{pronet,piamrat2011},
codec-switching \cite{mollericc2009}, codec bitrate manipulation \cite{jammehtelsys2012,Pan2013telsys},
and routing \cite{Tranrouting}. This paper addresses the challenge
of QoE provisioning in HANs using handoffs. A handoff is a process
of migrating from one network access point (AP) to another (belonging to the same of different wireless networks)\cite{Stemm1998}.
Handoffs are usually facilitated by mobility protocols such as Mobile
IPv4/v6 \cite{PerkinsOctober1996,mipv6}, Multi-homed Mobile IP \cite{ahlund2003},
and Stream Control Transmission Protocol (SCTP) \cite{sctp}. During
handoffs, a mobile node (MN) goes through several phases such as,
network discovery, network registration and network configuration.
These phases may cause severe degradation to user's QoE due to increase
in delay, jitter and packet losses \cite{marshgronvall,Bernaschi2007,mollericc2009}.
Since the late 1990's, much work has been done to optimize handoffs
in HANs (see, \cite{mobilitysurvey2010,wang2012}). However, there
is a dearth of research considering the challenge of QoE-aware handoffs
in HANs \cite{pronet,mobilitysurvey2010}. 

One of the major challenges for QoE-aware handoff is accurate QoE
 prediction for the available wireless networks \cite{pronet,mitraatnac}.
Further, HANs are prone to stochastic network conditions such as wireless network
congestion and wireless signal fading. Most of the approaches \cite{Balasubramaniam2004,nasser2006,m4,navarromdp}
to handoffs do not explore the benefits of QoE  prediction for QoE-aware
handoffs \cite{pronet,mitraatnac}. Few approaches \cite{piamrat2008,piamrat2011,varela2011,Rosario2013,Quadros2013},
consider QoE-aware handoffs, but assume the availability of underlying network path 
probing schemes to predict user's QoE \cite{varela2011,mitraatnac}.
This assumption is unrealistic and severely hinders QoE provisioning
in HANs \cite{mitraatnac}. Lastly, current methods \cite{m4,piamrat2011,varela2011,Rosario2013,Quadros2013}
for QoE-aware handoffs are myopic and inefficient i.e., these methods
do not consider learning about previous network conditions and actions
(regarding handoffs) taken by the MN for QoE provisioning,
therefore, limiting their ability to make efficient QoE-aware handoffs
\cite{pronet}. We assert that efficient QoE prediction and provisioning in
HANs remain an unresolved and a challenging problem \cite{pronet,mobilitysurvey2010,mitraatnac,qoeprovisioning2013}.

\subsection{Our Contributions }

In this paper, our contributions are two-fold. In particular, we present: 
\begin{enumerate}
\item A novel method for end-to-end QoE prediction using a passive
probing mechanism based on hidden Markov models (HMMs) \cite{Rabiner1989}
and Multi-homed Mobility Management Protocol (M-MIP) \cite{ahlund2003}.
The proposed method eliminates the need for additional probe packets
on network interfaces while predicting QoE states accurately. To the
best of our knowledge, the method proposed in this paper is the first
to integrate QoE prediction capabilities directly into a mobility
management protocol. Our results show that average QoE prediction
accuracy of 97\% is achieved considering WLAN and a cellular network.

\item A novel method for proactive QoE-aware handoffs based on
Reinforcement Learning (RL) \cite{russelandnorvig}. We compare our
method with Multimedia Mobility Manager $(M^{4})$ \cite{m4} and
a naive scheme (similar to \cite{Balasubramaniam2004,nasser2006})
and show that our proposed method outperforms them by reducing the
average number of handoffs by 60.65\%. 
\end{enumerate}
The rest of the paper is organized as follows. Section 2 presents
the related work. Section 3 presents PRONET, our approach for proactive
context-aware QoE provisioning in HANs.In presents 4 two
analytical models for QoE prediction and for proactive QoE-aware handoffs.
Section 4 presents our results based on extensive simulations and
real trace-based analysis. Finally, section 5 presents the conclusion
and future work.

\section{Related Work}

\subsection{Handoffs in Heterogeneous Access Networks}

Significant interest in the areas of vertical handoffs and HANs has
been shown in the past decade \cite{Stemm1998,Wang1999,Prehofer,Indulska2004,Balasubramaniam2004,McNair2004,Vidales2004,nasser2006,m4,piamrat2008,mollericc2009,mobilitysurvey2010,piamrat2011,wang2012}.
The term \emph{vertical handoff }came into existence in late 1990s
during the inception of The Bay Area Research Wireless Access Network
(BARWAN) project \cite{Katz1996}. The main aim of the BARWAN project
was to develop a wireless overlay network structure and to facilitate
vertical handoffs with minimum latency. An overlay network structure
consists of different overlays (or cells) of different sizes belonging
to different wireless network technologies such as Infrared, WaveLan
and Richonet. 

Stemm and Katz \cite{Stemm1998} proposed a system to facilitate vertical
handoffs in overlay networks. Their system considered handoffs between
Infrared, WaveLan and Richonet wireless network technologies. The
authors considered MIPv4 \cite{Perkinshttp:} protocol for mobility
management where the handoffs were triggered based on signal to noise
ratio (SNR) values. The authors proved that vertical handoffs between
different overlays were possible with handoff latencies in the range
of 181 ms to 2.53 sec, making roaming feasible. 

Wang \emph{et al.} \cite{Wang1999} presented a novel policy-based
system for vertical handoffs. Policies were based on cost, power and
network conditions. Their policy model computed costs based on parameters
such as bandwidth and power. The proposed model was flexible to a
certain degree, that a user had choices between the parameters to
facilitate user-centric handoffs i.e., he/she could choose bandwidth
over cost or vice versa. The main advantage of their model was that
it provided seamless connectivity without human intervention. Policies
were also clearly defined and could be modelled as per user requirements. 

McNair and Zhu \cite{McNair2004} presented the design and performance
issues for achieving adaptable vertical handoffs in a multi-network
4G environment. They discussed the possible architectural components
and handoff metrics that can be used in conjunction with the conventional
signal strength measurements to facilitate vertical handoffs. We classify
the aforementioned methods as \emph{traditional methods} as they mainly
consider layer 2 (of the OSI protocol stack \cite{Zimmermann1980})
parameters to trigger handoffs. These parameters include signal strength,
bit error rate and beacons. 

Since the later-half of 2003, context-aware methods were proposed by several
researchers \cite{Prehofer,Indulska2004,Vidales2004,nasser2006}.
\emph{Context} is any information that assists in determining a situation(s)
related to a user, network or device \cite{Dey}. In this paper, we
define c\emph{ontext as any information that assists in facilitating
QoE-aware handoffs}. Context can be static and dynamic. Static context
does not change often, while dynamic context changes over a period
of time and is difficult to predict. Static context may include user's
application preferences, their security requirements and cost. Dynamic
context may include user location, velocity, network load, battery
power, memory/CPU utilization, presence and SNR. Context can be collected
via sensors such as GPS for collecting user's $x$ and $y$ location
coordinates. It can also be collected via probes. For example, probe
packets sent between the two network entities can be used to determine
QoS statistics such as delay and packet loss \cite{SLMTELSYS2008}. The aim
of context-aware methods is to consider a combination of several parameters
to facilitate efficient handoffs. 

Prehofer and Wei \cite{Prehofer} presented a framework for context-aware
vertical handoffs. in particular, they  developed a flexible context exchange
protocol incorporating mobile agents for context collection.
However, the authors do not perform any experimental/simulation studies to
validate their proposed system.

Indulska and Balasubramanian \cite{Indulska2004} presented a context-aware
handoff solution for WLAN and GPRS networks. The authors divided context
into four categories i.e., \emph{sensed context}:\emph{ }such as user
location and QoS; \emph{static context}: such as device capability
in terms of CPU type, screen size and content capability; \emph{profiled
context: }such as personal settings and user cellular network profile;
and \emph{derived context}: such as network profile. The authors did
not consider mobility management protocols, instead they relied on
proxy agents on gateway nodes to relay network traffic between different
networks. The authors using experimentation validate that their method
can facilitate vertical handoffs. 

Vidales \emph{et al.} \cite{Vidales2004} presented a context-aware
and policy based solution called PROTON. However, their policies
were difficult to formulate and integrate in a real system. The authors
however, presented a thorough investigation of the effects of MIPv6
mobility management protocol on vertical handoffs.

Most of the context-aware methods are based on multi-attribute decision
making (MADM) \cite{madm} which utilizes several context attribute
values to determine a single scalar value. The network with a higher
scalar value is the target for a handoff. For example, Indulska and
Balasubramanian \cite{Indulska2004} used Analytic Hierarchy Process
(AHP) for handoffs between WLAN and 3G. Nasser and Hasswa and Hassanein
\cite{nasser2006} used Simple Additive Weighting (SAW) for handoffs
between WLAN and a cellular network. 

We assert that in HANs, context can be highly dynamic and stochastic
i.e., it can change in a very short period of time and is uncertain;
it can be imperfect; it can exhibit a range of temporal characteristics;
it can have several alternative representations; it can be interrelated;
it can be distributed; and it may not be available at a particular
time. The timely collection and processing of context may be crucial
as it may loose its accuracy. The aforementioned methods may not efficiently
deal with uncertain context as they simply utilize available context
attribute values at current time-stamp $t$ to facilitate a handoff.
\textit{In this paper, we argue for and develop context-aware methods consider
past and present context attribute values to learn about the best
time to make a QoE-aware handoff. }

Stevens-Navarro, Lin and Wong \cite{navarromdp} presented a Markov
Decision Process (MDP) based method to facilitate handoffs in HANs.
Their scheme learns from current QoS values and past decisions regarding
handoffs to select the best wireless network for a handoff at time
$t.$ The authors compared their results with MADM-based methods (e.g.,
SAW and TOPSIS) and show that their method achieves higher expected
total reward and lower expected number of handoffs. However, their
method do not facilitate QoE-aware handoffs and assume that the path
probing method is present to determine QoS values. This assumption
is quite strict and unrealistic. In reality, QoS parameters have to
be estimated by either using passive or active probing. The QoS parameter
values may also be provided by any third party to the MN, if present
on the network.

\subsection{QoE Provisioning in Heterogeneous Access Networks}

As discussed in section 1, QoE provisioning using handoffs necessitate
accurate path probing mechanisms such that a MN can predict user's
QoE. Hidden Markov models (HMMs) \cite{Rabiner1989} have been successfully
applied \cite{Salamatian,liu2003,taoandguerin2004,mitraatnac} to
model both wired and wireless network characteristics. For example,
Liu, Matta and Crovella \cite{liu2003} considered HMMs and a pair
of RTT probe packets to classify different types of losses in wireless
networks. However, their scheme was limited to classification of congestion
and wireless losses and did not consider users\textquoteright{} QoE. 

Tao and Guérin \cite{taoandguerin2004} estimated Internet path performance
using HMMs. Their approach was similar to \cite{Salamatian} but they
extended this work by looking into sensitivity analysis of the model.
They also predicted the application behaviour with different characteristics
such as varying packet losses. The authors considered an active probing
mechanism to estimate and predict packet losses using HMMs. They
show that their approach works well in case of isolated packet losses
but is not accurate in case of bursty packet losses. This 
approach was also limited to wired networks and did not consider stochastic
wireless network impairments such as signal fading,  network congestion
and handoffs. Finally, active probing utilized in their approach requires additional probe packets which
increases bandwidth and monitoring costs. Compared to \cite{taoandguerin2004},
in this paper, we develop a passive probing mechanism based on HMMs
and M-MIP to estimate and predict user's QoE HANs. Our scheme do not
require additional probe packets thereby minimizing network bandwidth
and monitoring costs.

Tao \emph{et al.} \cite{taoicnp}\emph{ }presented a methodology for
QoE-aware path switching in wired networks. Their QoE prediction method
was based on predicting the mean opinion score (MOS) score for all
network paths based on simple Markov predictors. The main assumptions
of the proposed method were: the frequency of path-switching was low,
the delay variation was low and the effects of delay and path-switching
could be mitigated by the playback mechanisms. These assumptions are
restrictive, and may not be suitable in case of HANs, where a user may
move quite suddenly and abruptly in HANs leading to delay variation, abrupt handoffs due to signal fading, and packet losses. All these factors can significantly hamper users' QoE  \cite{mitrawcnc}. 

Marsh and Grönvall \cite{marshgronvall} and Marsh, Grönvall and Hammer
\cite{marsh2006} presented a study on the effects of handoffs on
VoIP applications. The authors developed a PSTN-based VoIP test bed
to evaluate the performance of VoIP applications in IEEE 802.11b networks.
The authors evaluated VoIP call quality based on six parameters including
packet loss, delay, jitter, link layer retransmissions, transmission
rate and signal to noise ratio (SNR). The authors also developed a
network score as a function of SNR, bandwidth, delay, jitter and packet
losses which validate that all of these parameters affect users' QoE
in WLAN. However, the problem with their approach was that they do
not map QoS parameters to QoE and hence, conclusive results could
not be obtained from end users' perspective.

Mobisense Project \cite{moller2008,mollericc2009,lewcio2009} was
carried out at Deutsche Telecom, Germany to investigate the effects
of handoffs on users' QoE. The results from this project were based
on WiFi and HSDPA networks. The authors studied the impact of delay,
audio bandwidth, packet loss and codec-switchover on users' QoE. It
was concluded that packet loss is the most important parameter that
significantly degrades users' QoE followed by audio bandwidth. However,
the authors do not propose methods for QoE-aware handoffs in HANs.

Piamrat \emph{et al.} \cite{piamrat2008,piamrat2011} and Varela and
Laulajainen \cite{varela2011}, presented methods for QoE-aware handoffs
based on the Pseudo-Subjective Quality Assessment (PSQA) metric \cite{psqa}.
The PSQA metric maps parameters such as percentage of packet loss
and mean loss burst size to MOS based on Random Neural Networks (RNNs).
The problem with RNN is that it require a large number of training
samples for performing accurate prediction. This limits its ability
to learn continuously in an on-line and unsupervised manner. 

Rosario \textit{et al.} \cite{Rosario2013} and Quadros \textit{et al.} \cite{Quadros2013} presented a method for QoE-aware handoffs by focussing on video QoE.  The authors aimed at estimating QoE using neural networks. In \cite{Quadros2013}, they extended their scheme presented in \cite{Rosario2013} and proposed a handoff initiation method using the IEEE 802.21 Media Independent Handover scheme. However, as mentioned previously, neural network based mechanisms require large amounts of training data. Further, the authors do not consider methods for network path probing and mobility management. Further, these methods are myopic and inefficient as they do not consider learning about previous network conditions and actions (regarding handoffs) taken by the MN for QoE provisioning. Therefore, limiting their ability to make efficient QoE-aware handoffs.

Tabrizi, Farhadi and Cioffi \cite{Tabrizi2012} presented a Q-learning based method for handoff management in HANs. Their proposed scheme looks promising. However, as with the other state-of-the-art methods for QoE provisioning in HANs (e.g.,\cite{piamrat2008,piamrat2011,varela2011,Rosario2013,Quadros2013}), the authors do not present a generic path probing scheme and the integration of their mechanism with a mobility management system. These systems were mainly validated using simulation studies and using real world prototype systems, leading to methods that may not work well in reality and may hinder efficient QoE-aware handoffs in HANs.

\subsection{Research Challenges}

From the discussion presented in this section, we identify the following
two research challenges for efficient QoE provisioning in HANs.
\begin{itemize}
\item \textbf{QoE prediction in heterogeneous access networks:} The current
approaches \cite{m4,piamrat2008,elktob2010,varela2011,piamrat2011}
for QoE-aware handoffs are limited. These approaches lack the availability
of efficient QoE prediction mechanisms. Further, approaches such as
\cite{piamrat2008,varela2011,piamrat2011} assume the availability
of path probing mechanism to make QoE-aware handoffs. These assumptions
are unrealistic and can severely hinder QoE provisioning in HANs \cite{pronet,varela2011,mitraatnac}.
We assert that current approaches to QoE-aware handoffs lack efficient
network path probing mechanisms and methods for accurate QoE prediction
in HANs. Path monitoring is not a new concept. Tao \emph{et al. }\cite{taoandguerin2004}
and Liu, Matta and Crovella \cite{liu2003} proposed to use active
and passive probing to estimate network path quality. Åhlund and Zaslavsky
\cite{ahlund2003,SLMTELSYS2008} proposed the use of passive probing
mechanisms using mobility management protocols for estimating network
path quality for making handoffs. However, these approaches were either
used for wired network environments or do not consider QoE  prediction
in HANs \cite{mitraatnac}. Thus, there is a need to develop novel
methods for QoE prediction in HANs.
\item \textbf{Proactive QoE provisioning in heterogeneous access networks:
}The current approaches \cite{m4,piamrat2008,elktob2010,varela2011,piamrat2011}
do not consider proactive QoE-aware handoffs by learning from past
and possible future network conditions. Network conditions in HANs
are stochastic due to uncertain and time-varying characteristics of
the network medium. Thus, there is a need to develop techniques that
are resilient to such problems by learning from underlying network
conditions and past actions taken by a MN regarding handoffs such
that users' QoE is maximized \cite{pronet}.
\end{itemize}

Compared to the aforementioned state-of-the-art methods, in this paper, we propose, develop and validate novel methods for: i.) QoE prediction using hidden Markov models and M-MIP-based passive probing mechanism; and ii.) proactive QoE provisioning using a Reinforcement Learning mechanism. To the best of our knowledge, none of the methods in the state-of-the-art presents such a comprehensive treatment of the problem of proactive QoE provisioning in HANs.


\section{PRONET: An Approach for Proactive QoE-Aware Mobility Management in Heterogeneous Access Networks}
This section presents PRONET---$ $An approach for proactive context-Aware mobility
Management to facilitate QoE-aware handoffs (see Fig. 1.). It follows a cognitive networking approach where a mobile node is able to sense the current network conditions, and then plan, decide and act under these conditions to achieve end-to-end goals \cite{pronet,Facchini2013}.
\textcolor{black}{In this paper, PRONET aims to reduce the number of handoffs while
maximizing user's QoE. }We consider a scenario where a mobile node
(MN) incorporating PRONET roams in HANs.\textcolor{black}{{} The MN
collects context attributes values such as}\textcolor{darkred}{{} }\textcolor{black}{network
delay and predicts user's QoE states using hidden Markov models (HMMs)
\cite{Rabiner1989}. The predicted QoE states are used by the Q-learning
algorithm \cite{sutton98} to proactively select the best network
interface for handoff.} The challenges addressed by PRONET are written
as:
\begin{enumerate}
\item Maximize QoE prediction accuracy $(\Phi)$ for two wireless network
interfaces $(i=|2|;WLAN\,\&\, Cellular)$$.$ This can be written
as: 
\begin{equation}
max(\Phi)\:\forall i \in I
\end{equation}

\item For a given handoff function $(\Gamma(\bullet))$, maximize user's \emph{QoE}
and reduce the average number of handoffs $(\Delta)$ between two
wireless network interfaces $(i=|2|;WLAN\,\&\, Cellular)$. This can
be written as: 
\begin{equation}
max(QoE)
\end{equation}
 
\begin{equation}
min(\Delta)
\end{equation}

\end{enumerate}
For vertical handoff decision problem, $\Gamma(\bullet)$ can be written
as: 
\begin{equation}
\Gamma=w_{j}(QoE_{i})+(1-w_{j})(Cost_{i})
\end{equation}
 where $w_{j}$ represents the weights associated with each parameter,
\emph{QoE} and \emph{Cost }and $\sum_{j=0}^{N}w_{j}=1$\emph{. }Cost
can be monitory cost, signaling cost and handoff switching cost. 
\begin{figure}
\centering{}\includegraphics[scale=0.75]{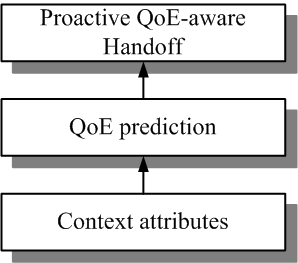}\caption{PRONET approach.}
\end{figure}

Both challenges necessitate accurate QoE prediction and provisioning
which can be achieved by developing accurate path probing mechanisms
\cite{pronet,varela2011} and by developing proactive methods for
QoE-aware handoffs in HANs \cite{pronet,mitraatnac}. The MNs can
either use active or passive network path probing for QoE prediction.
In case of active probing, extra probe packets are injected in the
network to mimic application traffic flow between two end-systems.
Based on delay and packet losses observed using these probes, network
states can be estimated, for example, ``network is congested'' or ``network is not congested''.
However, this technique while being beneficial, leads to an increase
in bandwidth requirements. Another probing technique is passive probing.
In this technique, the application traffic flow itself is used to
estimate path quality statistics without requiring additional probe
packets. Thus, bandwidth savings are made while estimating the path
quality. However, passive probing technique is application specific.
In this paper, \emph{we argue for application independent passive
probing techniques which can be used to monitor all network interfaces
of MNs, simultaneously}. In HANs, multi-homed mobility management
protocols such as M-MIP \cite{ahlund2005} are considered for application
session continuity using handoffs. These protocols implement signaling
mechanisms such as binding updates (BU) and binding acknowledgments
(BA) to handle events like packet flow redirection. Thus, in this
paper:
\begin{itemize}
\item We propose to use BU and BA packets as probe packets. Thus, eliminating
the need for additional probe packets generated on all the network
interfaces. Further, this method remains independent of any application
type. 
\item We propose to use HMMs trained using the one-way delay (OWD) or round
trip time delay (RTT) computed using the BA and BU packets to estimate
and predict QoE for VoIP applications.
\item We propose to use the Reinforcement Learning method (Q-learning) to facilitate
proactive handoffs that maximizes user's QoE over time while minimizing
the average number of vertical handoffs. 
\end{itemize}
Fig. 2 shows our high-level approach for network path quality prediction
using HMMs and M-MIP. A MN periodically exchange BU and BA messages
with the home agent (HA) or the correspondent node (CN). For each
BU sent by the MN to HA/CA, a corresponding BA is sent from HA/CN
to MN. The amount of time spent between sending a BA from HA/CN to
MN is the OWD. The amount of time spent between sending a BU from
MN and receiving a corresponding BA from HA/CN is the RTT delay. The
RTT delay is estimated at the network layer (considering the OSI model
\cite{Zimmermann1980}) of MN. By using RTT delay values, clock synchronization
is not required between the MN and HA as compared to OWD. The HMM
takes this OWD/RTT value as an input and try to predict the corresponding
QoE state at the application layer. This problem is not trivial as
the HMMs need to discover the hidden QoE states from the underlying
network conditions. 

Fig. 3 shows our method (represented as a dynamic
decision network) for proactive QoE-aware handoffs where we consider
Reinforcement Learning (RL) \cite{sutton98}.
In particular, the prediction results from the HMM are taken as inputs
by the RL algorithm to proactively select the best network interface,
$i \in I$ in terms of QoE. In this paper, we consider Q-learning \cite{sutton98}
as the RL method. Using Q-Learning,\emph{ }a MN learns an action-value
pair to select the best available network that maximizes user's QoE.
We will show later that our integrated approach (PRONET) minimizes the average number
of handoffs (which causes increase in packet loss and jitter \cite{marshgronvall,mollericc2009})
while maintaining high QoE levels. In the following sections, we present
two analytical models for QoE prediction and provisioning in HANs.

\subsection{Hidden Markov Models for QoE Prediction}

\begin{figure}
\centering{}\includegraphics[scale=0.25]{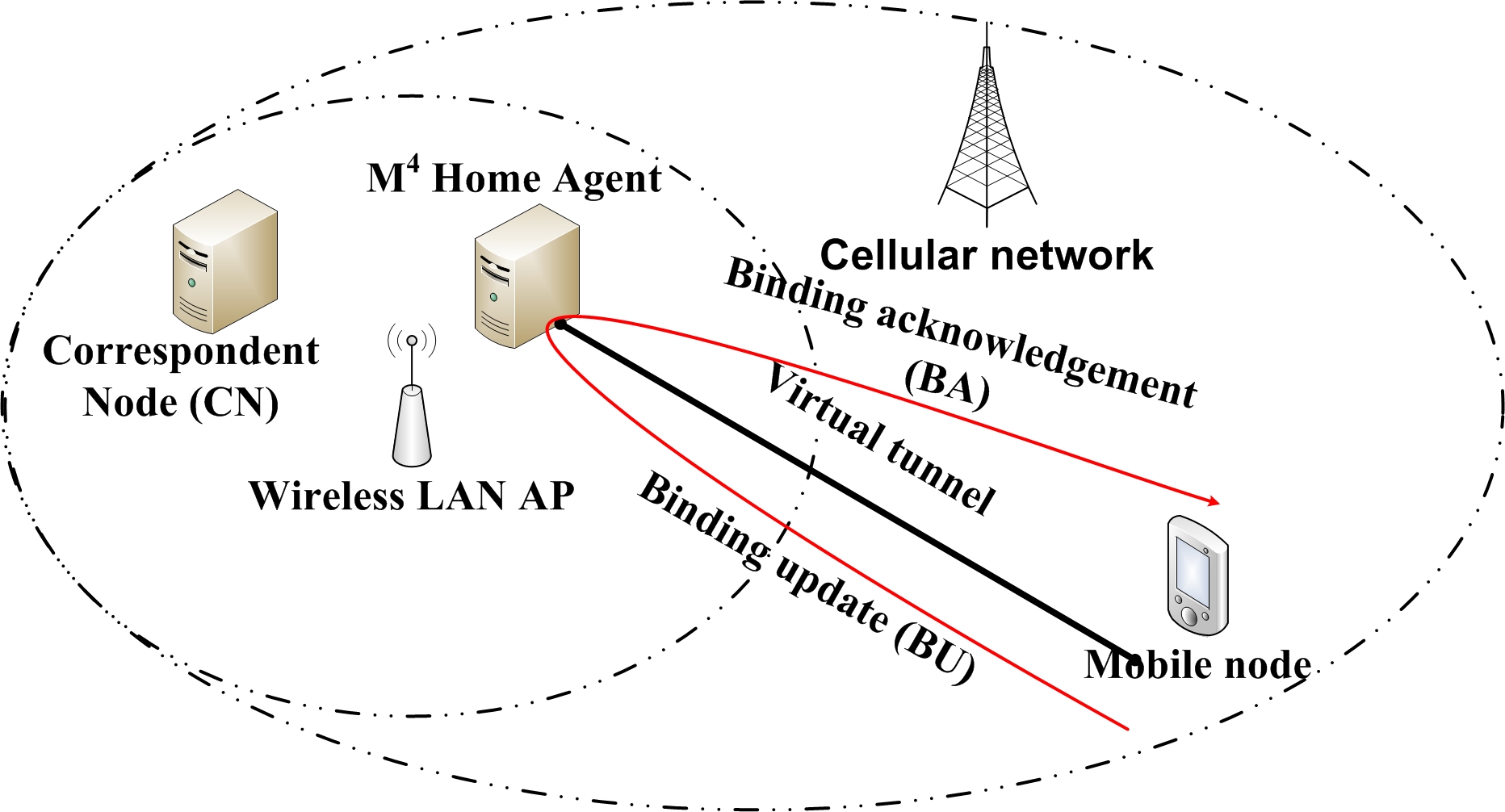}\caption{MN calculates one-way (OWD) or round-trip time (RTT) delay using BU
and BA messages. These delay values are then used by HMM for QoE  prediction.
The predicted QoE values are then used by RL algorithm for proactive
QoE adaption. }
\end{figure}

\begin{figure}
\centering{}\includegraphics[scale=0.4]{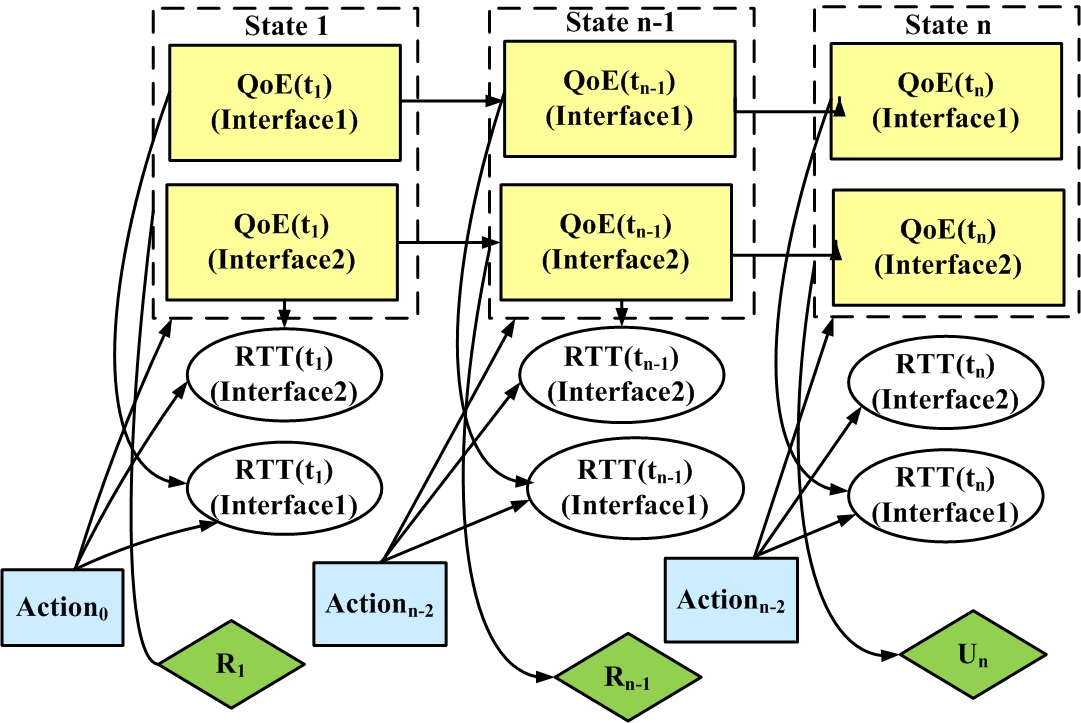}\caption{A dynamic decision network (DDN) representation of a QoE provisioning
scheme.}
\end{figure}

 We consider discrete-time HMMs \cite{Rabiner1989} for QoE prediction in HANs. HMMs are temporal probabilistic models in which a system state, in our case a QoE state, is described by a single discrete random variable. Fig. 4 shows our HMM for QoE prediction. 

\textbf{System states}: We use the notation $S_{QoE_{n}}^{t}$ to represent the QoE state at time $t$ where $t\in T$ and $T$ is the set of finite integers. At a given \emph{$t$,} our system can be in any state $n\in N$. For example, $|N|=5$. Thus, the state space can be written as \{$S_{QoE_{1}}^{t},S_{QoE_{2}}^{t},...,S_{QoE_{5}}^{t}$\}. When a system is in a particular state $S_{QoE_{n}}^{t}$, it outputs an observation or evidence ($E^{t}$). 

\textbf{Observations:} In our system, the evidence is the current \emph{OWD/RTT delay value } which is modelled as a Gaussian distribution in the form of: $P(E^{t}|S^{t}_{QoE})=\mathcal{N}[E^{t};\mu,\sigma^{2}]$, where $\mu$ is the mean and $\sigma^{2}$ is the variance. To describe state evolution over time, a QoE state transition matrix ($TM$) is defined. In this paper, we consider a first-order Markov process i.e., the current state is dependent only on previous state. It is shown in \cite{liu2003,taoandguerin2004} that first-order Markov process is sufficient for modelling temporal characteristics of a network channel (both wired and wireless). Thus, $TM$ is defined as: $P(S_{QoE}^{t}|S_{QoE}^{t-1})$. To start the process, an initial state distribution is defined. It is represented as $p=P(S_{QoE_{n}}^{t=0})$.

\textbf{Learning model parameters:} In HMMs, the problem of learning is that of learning the model parameters $\Theta$:$\{\mu,\sigma^{2},p,TM\}$. We consider the use of expectation maximization (EM) algorithm to train our HMMs. In EM algorithm, there are two main steps: E-step which computes posteriors over states and the M-step which adjusts the model parameters to maximize the likelihood of posteriors calculated in the E-step. It is an iterative process leading to a guaranteed increase in log-likelihood of the model ($log(\Theta)$) until convergence.

\textbf{State  prediction:} After learning the model parameters using EM, our main task is to perform QoE state prediction. It is a task of computing posterior distribution over the future states, given evidence till now. In this paper, we are interested in one-step QoE prediction.
\begin{figure}
\centering{}\includegraphics[scale=0.35]{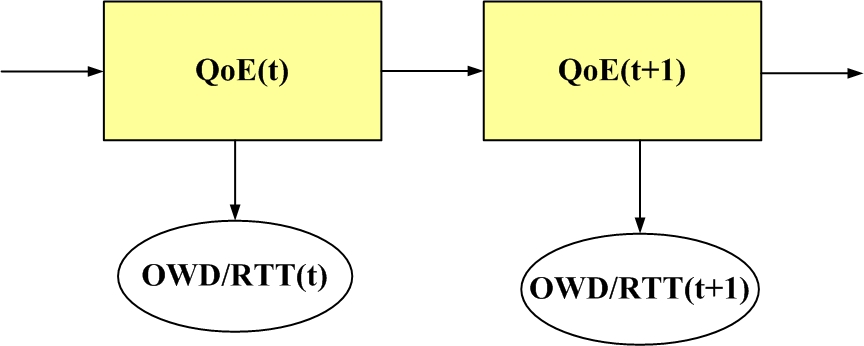}\caption{Hidden Markov model (HMM) for QoE  prediction using OWD or RTT delay
calculated using BA and BU of M-MIP.}
\end{figure}

\subsection{Reinforcement Learning for Proactive QoE-aware Handoffs}

The previous section considered the challenge of QoE prediction over
time in stochastic HANs conditions. In this section, we consider the
challenge of proactive QoE provisioning in HANs based on reinforcement
learning (RL) \cite{sutton98}. In this paper, we consider Q-learning
as the RL algorithm \cite{sutton98}. Using Q-Learning algorithm,
our \emph{agent}, the MN, learns an action-value pair to select the
best available network that maximizes user's QoE. The aim of our agent
is to predict QoE for all network interfaces $I$ and then proactively
select a network interface $i$ that maximize user's QoE; where $i\in I$
and $|I|=2\, and\, I=\{``WLAN"\, or\,``Cellular"\}$. This process of
learning and adaptation repeats until the system stops.

\subsubsection{Agent Design and System Model}

To proactively select a network $i \in I$, MDP is considered incorporating
a tuple $(TM,\Re,A,\pi,S,\gamma)$. We use $S$\footnote{We now use the term $S^{t}$ instead of $S_{QoE}^{t}$ to represent
QoE state at $t.$%
} and $TM$ from the previous section representing the state and the
transition model of the agent, respectively. $\gamma\in[0,1)$ is
called the \emph{discount factor}. Our agent, the MN, senses the \emph{environment}
(e.g., network conditions) and then using HMMs predicts the QoE states $\forall I$. Based on the predicted QoE state, it selects $i$ that
maximizes user's QoE in the long run. An agent and the environment
interacts with each other continually at fixed time-steps or \emph{decision
epochs} $(t=0,1,2,...,n;\, t\in T)$. $S_{i}^{t}$ and $S_{i+1}^{t}$
denotes the QoE state for all the network interfaces $i\,\in\, I.$
Further, $S_{i}^{t}$ and $S_{i+1}^{t}$ have their own independent
sensor models $E_{i}^{t}$ and $E_{i+1}^{t}.$ These models are learnt
independently using HMMs described in the previous section. 

\textbf{States: }Compared to \cite{navarromdp,Songmdp,jpsinghmdp},
we do not consider crude QoS metrics for state representation as it
can lead to an explosion in state space and increase in algorithm's
convergence time \cite{jpsinghmdp}. This is due to the fact that
the QoS values related to delay, jitter, bandwidth and packet loss
rate are continuous. Thus, to represent them as individual states,
they have to be quantized into finite bins of size $|N_{n=1}^{\infty}|$.
This way, the complexity grows quickly as the state space will be
the Cartesian product of all states related to each QoS parameter
for each network interface. It can be expressed as: $I\,\times S_{Delay(|N|)}^{I}\,\times S_{jitter(|N|)}^{I}\,\times S_{packetloss(|N|)}^{I}\,\times S_{bandwidth(|N|)}^{I}$;
where \emph{I} is the number of network interfaces. Further, in reality,
quantization of the state space may lead to a decrease in prediction
accuracy \cite{russelandnorvig}. 

To alleviate this problem, in this paper, we propose to use\emph{ finite
QoE states learnt using HMMs}. \textcolor{black}{For example, MOS
values in the range of 1 and 1.9 denotes state 1; MOS values in the
range of 2 and 3.5 denotes state 2 and MOS values in the range of
3.51 and 5 denotes state 3.} It
is important to note that our HMMs will automatically discover these
QoE states based on the training data. This way, the complexity of
$I\,\times S_{Delay(|N|)}^{I}\,\times S_{jitter(|N|)}^{I}\,\times S_{packetloss(|N|)}^{I}\,\times S_{bandwidth(|N|)}^{I}$
is substantially reduced to just $I\,\times\, S_{QoS(|2\, to\,5|)}^{I}$.
In this paper, we assume that QoE is computed using the ITU-T E-Model
\cite{G.107}. However, QoE can be computed using any methods presented
in the state-of-the-art \cite{mitraqoereview14}. To determine complex
QoE metric such as the ITU-T E-Model \cite{G.107}, all parameters
of the model should be known. In this paper, we show that only one
parameter (e.g., RTT delay/OWD) can be used to predict QoE states
using HMMs, eliminating the need for all additional parameter values. 

\textbf{Actions:} At each $t,$ an agent observes the environment
state(s), $S^{t}.$ Based on the observed state, an agent selects
an \emph{action, $a_{t}\in\text{A}$}: $A$ denotes the set of actions,
where $|A|=2;\, and\, A=\{``select\, i" \, or\,``select\, i+1"\}$. 

\textbf{Rewards:} After selecting an action \emph{$a_{t}$, }the agent
receives a numerical \emph{reward}, $r^{t+1}\in\Re(S)$ where $\Re:S\mapsto\mathbb{R}$
is the \emph{reward function}. After a reward is received, the agent
then transitions to the new state represented as $S^{t+1}.$

\textbf{Policy:}\emph{ }The mapping of a state to action $(\pi:\, S\mapsto A)$
is called a \emph{policy, $\pi.$ }In a stochastic environment, when
each policy is executed, it leads to a different environment history.
Thus, the quality of policy is determined by considering total rewards
it receives at each $t$ in the long run.

The reward function reflects QoE and cost associated with the chosen
$i.$ It is important to understand that in RL problems, positive
rewards (e.g.,\emph{ }related to\emph{ QoE}) are usually given to
the agent for every correct action it takes. On the other hand, negative
rewards (e.g.,\emph{ }related to\emph{ cost}) are given to the agent
when it makes a wrong decision. Costs can be related to signaling,
monitory budget or both. We use Eq. 4 to determine our reward function. Therefore,
$\Re(s)$ can be substituted for $\Gamma(\bullet)$. Thus, we now write the reward as:

\begin{equation}
\Re(s)=w_{j}(QoE_{t})+(1-w_{j})(Cost_{t})
\end{equation}

where $f(QoE)$ and $f(cost)$ can be solved by the following equations:

\begin{equation}
f(QoE)=\begin{cases}
1, & if\, QoE_{t}\geq QoE_{max}\\
\frac{QoE_{t}-QoE_{min}}{QoE_{max}-QoE_{min}}, & if\, QoE_{min}<QoE_{t}<QoE_{max}\\
0, & if\, QoE_{t}\leq QoE_{min}
\end{cases}
\end{equation}

\begin{equation}
f(Cost)=\begin{cases}
1, & if\,0<\, Cost_{t}\leq Cost_{min}\\
\frac{Cost_{max}-Cost{}_{t}}{Cost_{max}-Cost_{min}}, & if\, Cost_{min}<Cost_{t}<Cost_{max}\\
0, & if\, Cost_{t}\geq Cost_{max}
\end{cases}
\end{equation}
where $QoE_{t}$ represents the QoE state value at time $t.$ Similarly,
$Cost_{t}$ represents the cost at time $t$ and $w_{j}$ represents
the weights associated with each parameter, \emph{$QoE_{t}$} and
\emph{$Cost_{t}$ }and $\sum_{j=0}^{N}w_{j}=1$. 

The aim of our agent is to choose actions over time to maximize the
total payoff $ $or utility $(U)$. The agent executes some policy$(\pi)$
when it is in some state $S$ i.e., an agent takes an action $a=\pi(S).$
Thus, we write $U^{\pi}(S)$ as: 
\begin{equation}
U^{\pi}(S)=E\left[\sum_{t=0}^{\infty}\gamma^{t}\Re(S^{t})|\pi,S_{0}=S\right]
\end{equation}

The utility function $U(S)$ enables the agent to select actions \emph{$(a)$}
that maximize the expected utility. The next task is to find an optimal policy represented as 
$\pi^{*}$ at each time $t$ that $argmax(\Re(s)).$ It can be denoted as:
\begin{equation}
\pi^{*}=argmax_{\pi}E\left[\sum_{t=0}^{\infty}\gamma^{t}\Re(S^{t})|\pi\right]
\end{equation}

The utility function $U(S)$ lets the agent select actions using the
maximum utility principle i.e., an agent chooses actions that maximizes
the expected utility of the next state $S^{t+1}$. It is denoted as:

\begin{equation}
\pi^{*}(S)=argmax_{a}\sum_{S^{t+1}}TM(S,a,S^{t+1})U(S^{^{t+1}})
\end{equation}

From \cite{russelandnorvig}, we note that the utility of the state
is the immediate reward $\Re(S)$ plus the discounted utility of next
state $S^{t+1}$, provided an optimal action is chosen. Here $TM$ is the transition matrix as mentioned in section 3.1. Thus, the
utility of the state can be written as:

\begin{equation}
U(S)=\Re(S^{t})+\gamma\, max_{a}\sum_{S^{t+1}}TM(S,a,S^{t+1})U(S^{^{t+1}})
\end{equation}
This equation is the well-known Bellman equation \cite{russelandnorvig}
where $\Re(S^{t})$ is the immediate reward which is received by starting
from the current state; $\gamma\, max_{a}\sum_{s^{t+1}}TM(S,a,S^{t+1})U(S^{t+1})$
is the expected sum of discounted rewards by starting from the state
$S^{t+1}$. $S^{t+1}$ is distributed based on $TM(S,a,S^{^{t+1}})$
which is the distribution over where the agent will end up after executing
the first action $ $$a$. Bellman equations are used to solve $U(S)$
efficiently. Bellman equation can be solved using the value and policy
iteration algorithms \cite{russelandnorvig,sutton98}. However, it
requires that the $TM(S,a,S^{t+1})$ is known which might be hard
to estimate in stochastic network conditions in HANs. Thus, we consider
the Reinforcement Learning techniques to learn and select optimal
actions under uncertainty. In particular, we consider the Q-learning algorithm \cite{sutton98}.

\subsubsection{Q-learning for Proactive QoE-aware Handoffs}

We consider M-MIP enabled MN (Q-learning agent) that roams in HANs.
The design of our agent is given in Fig. 5. At each time $t,$ the agent
observes a QoE state (predicted using HMMs) and receives a reward. It then
learns and selects an action $a$ that maximizes the reward at the
next state $S^{t+1}$ at time $t+1.$ In Q-learning, the Q-value or $Q(S,a)$
is maintained in the Q-table of  size $|S|\times|A|.$ The immediate
reward $\Re(S)$ is the maximum reward the agent receives at $t+1$
by performing $a$ at $t.$ For each $(S,a)$, an action $a$ is either
rewarded or punished. For example, $Q(S_{(i=WLAN)}^{t} = ``QoE\,is\,excellent",a=$\emph{``do
not handoff to HSDPA''}) will be rewarded and $Q(S_{(i=WLAN)}^{t}= ``QoE\,is\,excellent",a=$\emph{``handoff
to HSDPA''}) will be punished. This is because, in the first case,
QoE is\emph{ {}``excellent''} and there is no need to make a handoff.
In the second case, the agent should learn via punishment that if
QoE is \emph{{}``excellent''}, handoffs should not occur. In case
of rewards, the Q-values are increased. In case of punishments, the
Q-values are decreased. This process of learning the $state-action$
continues till the goal state (or system stops) is reached. Thus,
the expected reward $E[\bullet]$ is the sum of discounted rewards
an agent collects in the long run. Thus, the value of taking an action
$a$ in state $S$ given a policy $\pi$ is denoted as $Q^{\pi}(S,a).$
It determines the expected return starting from $S$, taking an action
$a$ and then following a policy $\pi.$ It can be written as:
\begin{figure}
\centering{}\includegraphics[scale=0.45]{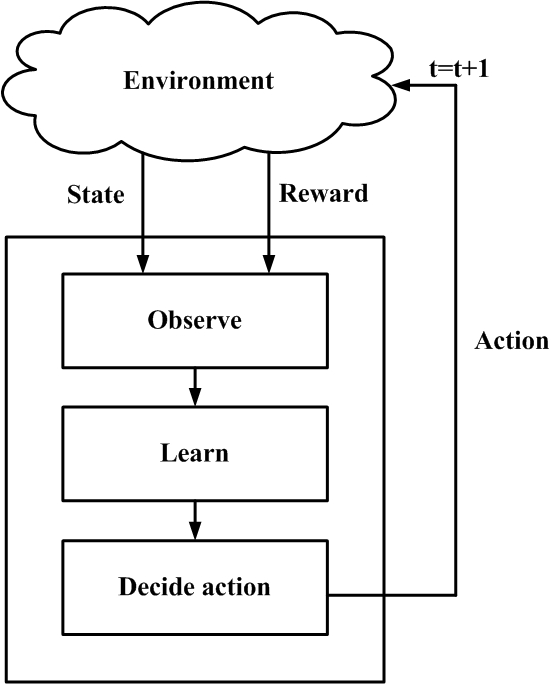}\caption{A Q-learning agent design for the mobile node.}
\end{figure}

\begin{algorithm}
\caption{\textbf{ An Algorithm for Proactive QoE-aware Handoffs.}}

\texttt{\textbf{Inputs:}}\texttt{ State $(S)$, reward signal $(\Re)$}

\texttt{\textbf{Output:}}\texttt{\emph{ }}\texttt{An action, $a \in A$}

\texttt{\textbf{Initialize}}

\texttt{\textbf{$\quad$$\quad$$\quad$$\quad$}}\texttt{$\quad$$\quad$$\quad$
$S$, $a$, $Q(S,a)\leftarrow0$, $\Re\leftarrow0$}

\texttt{\textbf{Run}}

\texttt{\textbf{$\quad$}}\texttt{/* Network discovery */}

\texttt{\textbf{$\quad\quad\quad\quad\quad$}}\texttt{1. Using RSSI,
discover $i\,\in\, I$}

\texttt{\textbf{$\quad$$\quad$For each}}\texttt{ $i\, \in I$}

\texttt{\textbf{$\quad$$\quad$$\quad$$\quad$if}}\texttt{ RSSI$(i)$
$\geq$ threshold(RSSI$(i)$)}

\texttt{\textbf{$\quad$$\quad$$\quad$$\quad\quad$}}\texttt{2. Connect
to $i$}

\texttt{\textbf{$\quad$$\quad$$\quad$$\quad$End if}}

\texttt{\textbf{$\quad$$\quad$End for}}\texttt{ }

\texttt{\textbf{$\quad$$\quad$}}\texttt{/* Network configuration */}

\texttt{\textbf{$\quad$$\quad$For each}}\texttt{ $i\, \in I$}

\texttt{\textbf{$\quad$$\quad$$\quad$$\quad\quad$}}\texttt{3. Establish
tunnel with HA}

\texttt{\textbf{$\quad$$\quad$End for}}

\texttt{\textbf{$\quad$}}\texttt{ /* Proactive QoE-aware handoff */}

\texttt{\textbf{$\quad$$\quad$if}}\texttt{ ``exploitation mode''$==$ ``true''}

\texttt{\textbf{$\quad$$\quad$$\quad$$\quad$$\quad$}}\texttt{4. Predict
the state $S$ using HMM}

\texttt{\textbf{$\quad$$\quad$$\quad$$\quad\quad$}}\texttt{5. Select $a$ randomly and execute it}

\texttt{\textbf{$\quad$$\quad$$\quad$$\quad$$\quad$}}\texttt{6. Go
to 4 and repeat till the end of session}

\texttt{\textbf{$\quad$$\quad$Else if}}

\texttt{\textbf{$\quad$$\quad$For each}}\texttt{ $i\ \in I$}

\texttt{\textbf{$\quad$$\quad$$\quad$$\quad$$\quad$}}\texttt{
/* Exploration mode */}

\texttt{\textbf{$\quad$$\quad$$\quad$$\quad$$\quad$}}\texttt{7. Predict
the state $S$ using HMM}

\texttt{\textbf{$\quad$$\quad$$\quad$$\quad$$\quad$}}\texttt{8. Select
$ $$a$ with $maxQ(S,a)$ and execute it}

\texttt{\textbf{$\quad$$\quad$$\quad$$\quad$$\quad$}}\texttt{9. Receive
$\Re$}

\texttt{\textbf{$\quad$$\quad$$\quad$$\quad$$\quad$}}\texttt{10. Update
the $Q(S,a)$ using Eq. 15}

\texttt{\textbf{$\quad$$\quad$$\quad$$\quad$$\quad$}}\texttt{11. Go
to 6 and repeat till the end of session}

\texttt{\textbf{$\quad$$\quad$}}\texttt{ \textbf{End if}}

\texttt{\textbf{$\quad$$\quad$End for$\quad$$\quad$$\quad$$\quad$$\quad$$\quad$}}

\texttt{\textbf{End}}

\end{algorithm}
 
\begin{equation}
Q^{\pi}(S,a)=E\left[\sum_{t=0}^{\infty}\gamma^{t}R(S^{t})|S_{0}=S,a_{0}=a,\pi\right]
\end{equation}
$Q^{\pi}$ is known as the \emph{action-value} function for $\pi.$
The optimal policy $\pi$ shares the same action-value function, $Q^{*}$
Thus, the optimal state-value function can be written as: 
\begin{equation}
Q^{*}(S,a)=max_{\pi}Q^{\pi}(S,a)
\end{equation}
i.e., for all states $S$, and all actions $a\in A$, the state-action
pair $(S,a)$ provides expected return for taking $a$ in $S$ and
then following the optimal $\pi.$ Thus, we can write $Q(S,a)$ in
terms of $U(S)$ (Eq. 11). It can be written as: 
\begin{equation}
U(S)=max_{a}Q(S,a)
\end{equation}

The aim of $Q(S,a)$ is to learn the action value function.
Q-learning is a model-free RL approach and does not require $TM(S,a,S^{t+1})$
and can directly relate to the utility values. It can be solved recursively
using the following update equation:

\begin{equation}
Q(S,a)\leftarrow Q(S,a)+\alpha[\Re(S)+\gamma\, max_{a^{t+1}}Q(S^{t+1},a^{t+1})-Q(S,a)].
\end{equation}
$\alpha$ is the \emph{learning rate} which determines how much time
the agent should spend in learning the policies. This equation is
calculated whenever an action $a$ is executed in state $S^{t}$ leading
to state $S^{t+1}$. It has been proved when this equation is executed
infinite number of times, $Q^{t}$ converges to $Q^{*}$ with probability
one where the learning rate $\alpha$ decreases to 0 \cite{sutton98}.
Note that the agent can choose to be in either \emph{exploration mode}
or in the \emph{exploitation mode}. In the exploration mode, the agent
will select a random action to pursue learning. However, in the exploitation
mode, the agent will automatically select an action $a$ corresponding
to a state $S$ which has the maximum value in the Q-table. Algorithm
1 presents the algorithm for proactive QoE-aware handoffs in HANs.
 In this paper, we assumed two networks i.e., WLAN and cellular.\textcolor{darkred}{{}
}\textcolor{black}{In case more networks are to be incorporated (say,
LTE), another HMM for LTE interface can be learnt based on the BU/BA
messages received for that network interface. On the other hand, in
the Q-learning algorithm, the Q-table will have can also be easily
extended by adding the corresponding states for the LTE interface. }

\section{Results Validation}

\subsection{QoE Prediction in WLAN Using One-way Delay}

In this section, we use our HMM-based method to estimate and predict
QoE states based on\emph{ passive one-way delay} (OWD). The OWD was
calculated at the mobile node (MN) using binding acknowledgment (BA)
packets sent from the home agent (HA)/correspondent node (CN) to MN.

\subsubsection{Simulation Setup }

We performed several simulation studies using OPNET{\texttrademark}
network simulator \cite{opnet} and considered cases such as wireless
network congestion. Our simulation scenario is shown in Fig. 6. Table
2 shows the parameters and their values used for performing simulations
in case of wireless network congestion. We considered a MN, HA, CN
and three additional wireless nodes (WNs). To saturate the IEEE 802.11b
access point (AP), we set up the WNs to generate additional background
UDP traffic. The maximum achievable bit rate was approximately 5Mb/s
after which the AP dropped all packets due to buffer overflow. A CN
generated an additional 5 UDP packets (to mimic probe packets similar
to BA) per second to help the MN calculate path delay and QoE statistics.
Based on \cite{andersson2008}, we selected the size of each BA packet
as 24 bytes, generating 960 bps as the M-MIP overhead. 

During simulations, once the network reached its steady state, the
MN initiated a voice call to the CN and calculated OWD and QoE using
the probe packets. From simulation studies, we concluded that in case
of heavy UDP traffic, wireless network congestion occurred due to
increased end-to-end delay which in turn caused ITU-T E-Model \cite{G.107}
mean opinion score (MOS) to vary. In case of ITU-T G.711 voice codec,
the average MOS was 2.06. On the other hand, in case of ITU-T G.729
voice codec, the average MOS was 2.41. The collected time series of
OWD and the MOS were used to train the HMMs. 
\begin{figure}
\begin{centering}
\includegraphics[scale=0.25]{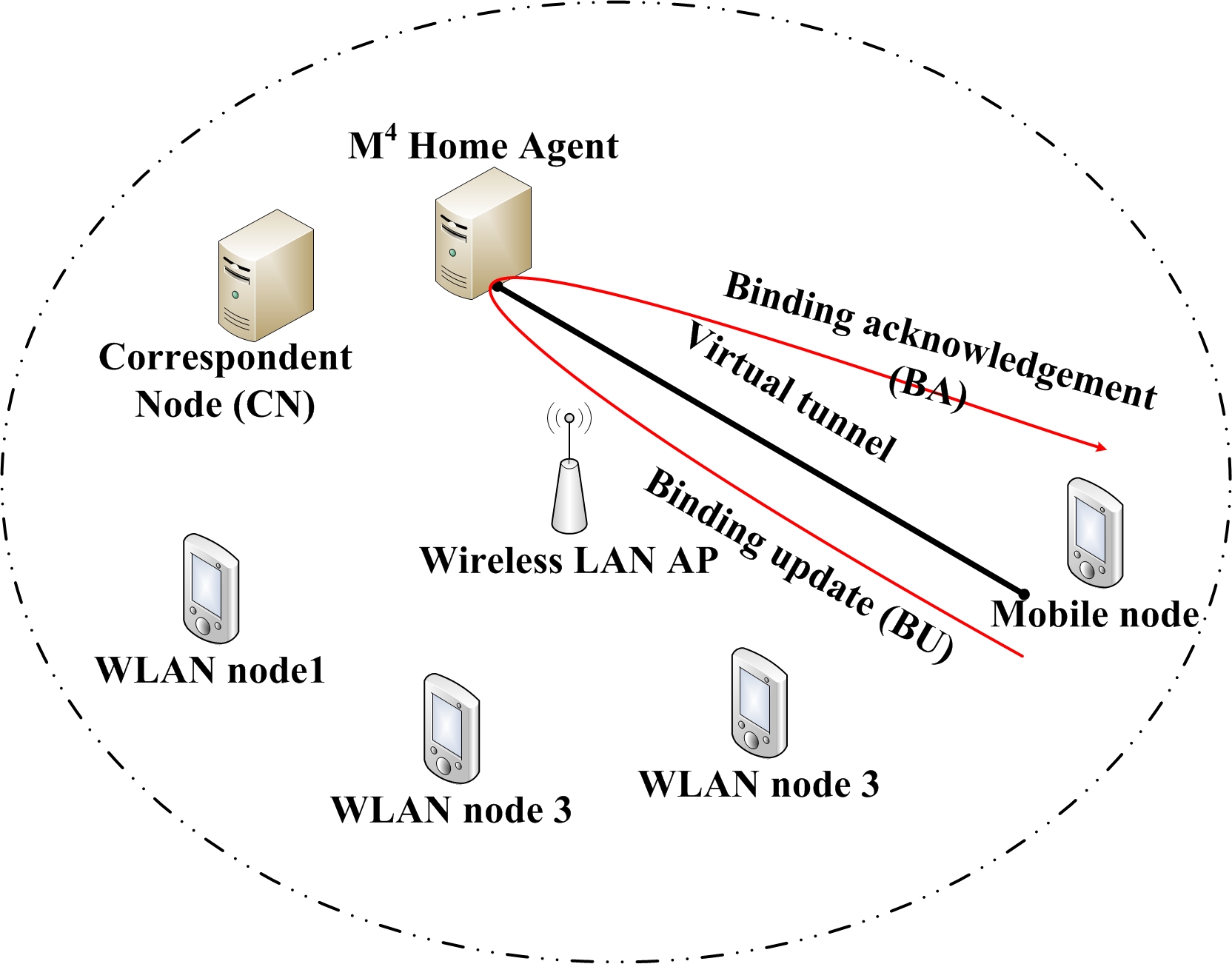}\caption{Wireless network congestion scenario.}

\par\end{centering}

\end{figure}
\begin{table}
\caption{Simulation setup parameters.}

\centering{}%
\begin{tabular}{|c|c|}
\hline 
\textbf{Parameter} & \textbf{Values}\tabularnewline
\hline 
\hline 
Codec & ITU-T G.711 \& ITU-T G.729\tabularnewline
\hline 
Network technology & IEEE 802.11b\tabularnewline
\hline 
No. of probe packets & 2,5,10\tabularnewline
\hline 
Size of the probe packet & 24 bytes \& 48 bytes\tabularnewline
\hline 
Movement pattern & stationary\tabularnewline
\hline 
Simulation time & 16 minutes\tabularnewline
\hline 
No. of simulations & 100\tabularnewline
\hline 
\end{tabular}
\end{table}

\subsubsection{Numerical Analysis}

To validate our proposed method, we initially considered 2 BA packets
(as probe packets) per second to calculate the OWD. However, the HMMs
could not efficiently predict QoE states on one second basis due to
missing BA values caused by bursty packet losses. Further, it may
happen that BA packets arrive too late (RTT/OWD>$t_{threshold}$),
where for example, $t_{threshold}=0.650\, sec$ in case of ITU-T G.729
codec \cite{G.107,G.113}. This led us to study the impact of 5 and
10 BA packets per second for calculating the OWD and predicting user's
QoE. From the simulation studies, we concluded that 5 UDP packets
per second are sufficient for QoE prediction even in case of high
and bursty packet loss conditions. 

To predict QoE states, we collected data using 100 simulation runs
(with different random seeds) and analyzed results related to both
ITU-T G.711 and ITU-T G.729 voice codecs. Similar to \cite{Iannelloetal2005},
we randomly selected 10 files out of 100 for each codec to train the
HMMs. Each file consisted of a 101 second time-series of OWD and the
corresponding QoE values. We used BayesNet Toolbox for MATLAB \cite{BNT}
for model parameter learning using EM algorithm and QoE state prediction.
One of the best approaches for evaluating a model's prediction accuracy
is to perform cross-validation \cite{russelandnorvig}. In cross-validation,
some fraction of the data is kept for training and the remaining data
is used as the test data. The training data and the test data are
randomly chosen for each fold. For our model validation, we considered
2- and 10-fold cross-validation.

We trained the HMMs based on three states. State 1 corresponds to
QoE values less than 2. State 2 corresponds to QoE values greater
than or equal to 2 and less than 3. Finally, state 3 corresponds to
QoE values greater than or equal to 3. For the sake of brevity, table
3 shows the learnt model parameters for ITU-T G.711 voice codec in
case of wireless network network congestion. The prior matrix $(p)$
and the transition matrix $(TM)$ were estimated as follows: \\

$\rho_{WLAN(G711)}^{congestion}=\begin{bmatrix}0.6000 & 0.2000 & 0.2000\end{bmatrix}$,\\

$TM_{WLAN(G711)}^{congestion}=\begin{bmatrix}0.9279 & 0.0596 & 0.0125\\
0.2817 & 0.3803 & 0.3380\\
0.0400 & 0.2400 & 0.7200
\end{bmatrix}.$\\

\begin{table}
\caption{HMM parameters learnt for ITU-T G.711 codec in case of WLAN network
congestion.}

\begin{centering}
\begin{tabular}{|c|c|c|c|}
\hline 
 & State 1 & State 2 & State 3\tabularnewline
\hline 
\hline 
mean $(mu)$ & 0.4850 & 0.1302 & 0.0462\tabularnewline
\hline 
var $(\sigma^{2})$ & 0.0576 & 0.0010 & 0.0006\tabularnewline
\hline 
\end{tabular}
\par\end{centering}

\end{table}

From the learnt transition matrix $(TM_{WLAN(G711)}^{congestion})$
in case of ITU-T G711 codec, we conclude that if the QoE on the path
is either state 1 or 3, it is likely that the QoE states will remain
constant for some time. If the QoE on path is in state 2, it is likely
that QoE will fluctuate between state 1 and state 3. \textcolor{black}{Fig
7. shows our model's prediction accuracy in case of WLAN congestion.
The average prediction accuracy was approximately 94\% considering
both ITU-T G.711 and ITU-T G.729 voice codecs.  Fig 8. shows an example of how HMMs can predict QoE states in complex network conditions such as WLAN network congestion. These results clearly validate that HMMs can accurately predict QoE in case of WLAN network congestion.We conclude that our scheme performs very well in case of both isolated as well as bursty packet losses, and in cases where OWD fluctuates significantly. }

\begin{figure} \centering{}\includegraphics[scale=0.20]{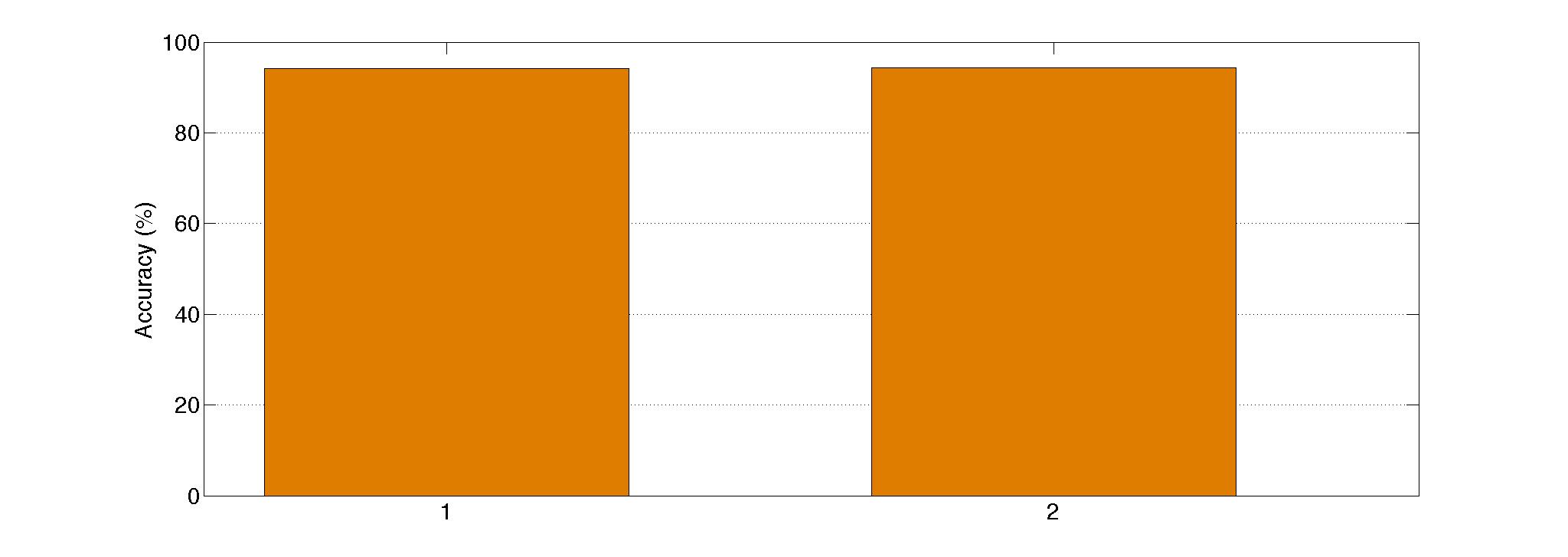}\caption{Prediction accuracy of HMM in case of WLAN congestion. 1.) ITU-T G.711 codec. 2.) ITU-T G.729 codec.} \end{figure}

\subsection{QoE  Prediction Using Round Trip Time Delay}

In this section, we use the BA/BU packet pairs belonging to M-MIP
\cite{ahlund2005} for passive RTT measurements. These RTT measurements
are used by HMMs to predict the QoE states for WLAN and CDMA 2000
networks.

\subsubsection{Experimental Setup}

\textcolor{black}{Fig. 2 shows our targeted scenario. }A multi-homed
MN, can roam between WLAN and cellular networks. The MN keeps on learning
and predicting the QoE states by using the RTT values computed using
the BU/BA packets exchanged between the MN and HA/CN. For results
validation, we obtained an experimental data set from the authors
in \cite{m4}. This data set contains 12 different time series of
RTT delay values for both WLAN and CDMA 2000 network interfaces corresponding
to 12 different experimental runs. These experiments were conducted
to understand the effects of roaming and handoffs on VoIP applications.
The RTT values were calculated at 1 second time interval while the
MN was on-the-move at different speeds (20 km/h and 30 km/h). Note
that the ITU-T E-Model \cite{G.107} uses OWD values to estimate MOS.
Therefore, in this data set, we used the RTT measurements and divided
them by 2 to derive the corresponding QoE values. We then used the
RTT and QoE values to train HMMs. The trained HMMs were used to predict
the QoE. As mentioned previously, the parameters for HMMs were learnt
using the EM algorithm in MATLAB. The model validation was done based
on 2-fold cross validation.

\subsubsection{Results Analysis}

We considered 2 and 3-state HMMs for QoE learning and prediction.
State 1 corresponds to QoE values less than 2. State 2 corresponds
to QoE values greater than or equal to 2 and less than 4. Finally,
state 3 corresponds to QoE values greater than or equal to 4. In case
of WLAN, our HMMs sufficiently learnt using only two states, i.e.,
using state 1 and state 3. There were very few state 2 values and
we folded them to state 3. This did not lead to the loss of prediction
accuracy. In case of CDMA 2000 network, there was sufficient data
related to all three states which was used for learning the model
parameters. Table 4 shows the learnt model parameters for ITU-T G.729
voice codec in case of WLAN and CDMA 2000 network interfaces. The
estimated transition matrix and prior are as follows: 

$\rho_{WLAN}^{G729}=\begin{bmatrix}0 & 1\end{bmatrix},$$TM_{WLAN}^{G729}=\begin{bmatrix}0.9500 & 0.0500\\
0.0654 & 0.9346
\end{bmatrix}$\\

$\rho_{CDMA2000}^{G729}=\begin{bmatrix}0 & 0 & 1\end{bmatrix},$ and
$TM_{CDMA2000}^{G729}=\begin{bmatrix}0.7852 & 0.1333 & 0.0815\\
0.1111 & 0.8148 & 0.0741\\
0.0696 & 0.0435 & 0.8870
\end{bmatrix}.$\\

From the learnt model parameters, and in case of WLAN, we conclude
that once the QoE is in state 1 or 3, it remains stable with probabilities
greater than 90\%. This suggests that WLAN either provides excellent
QoE (state 3) or performs poorly (state 1). The poor performance was
attributed to the MN moving out of the coverage area of WLAN and connecting
to CDMA2000 network interface. In case of CDMA2000 network interface,
once the QoE is in a particular state, it is likely that it will not
fluctuate much. However, in case of the ITU-T G.711 voice codec, if
the QoE is in state 2, there is a high probability that it might switch
to either state 1 or 3. 
\begin{table}
\caption{HMM parameters learnt for ITU-T G.729 codec in case of roaming.}

\begin{centering}
\subfloat[WLAN network interface]{

\centering{}%
\begin{tabular}{|c|c|c|}
\hline 
 & State 1 & State 2\tabularnewline
\hline 
\hline 
mean $(mu)$ & 0.9905 &  0.0519 \tabularnewline
\hline 
var $(\sigma^{2})$ & 0.0044 &  0.0079 \tabularnewline
\hline 
\end{tabular}}
\par\end{centering}

\centering{}\subfloat[CDMA 2000 network interface.]{

\centering{}%
\begin{tabular}{|c|c|c|c|}
\hline 
 & State 1 & State 2 & State 3\tabularnewline
\hline 
\hline 
mean $(mu)$ & 0.9519 & 0.6401 & 0.2857\tabularnewline
\hline 
var $(\sigma^{2})$ & 0.0055 & 0.0076 & 0.0025\tabularnewline
\hline 
\end{tabular}}
\end{table}

Fig 9 shows the prediction accuracy of the proposed method for two
voice codecs and for both WLAN and CDMA2000 network interfaces. Fig.
10 shows the capability of HMMs for QoE prediction for CDMA 2000 network
interface. The prediction accuracy was 100\% for WLAN and the prediction
accuracy for CDMA2000 was 95.60\%. Thus, the average prediction accuracy
was 97.80\%. From experimental analysis, we conclude that HMMs can
model complex time-series pertaining to QoE in realistic network settings
and are extremely beneficial for accurate QoE prediction.
\begin{figure}
\centering{}\includegraphics[scale=0.22]{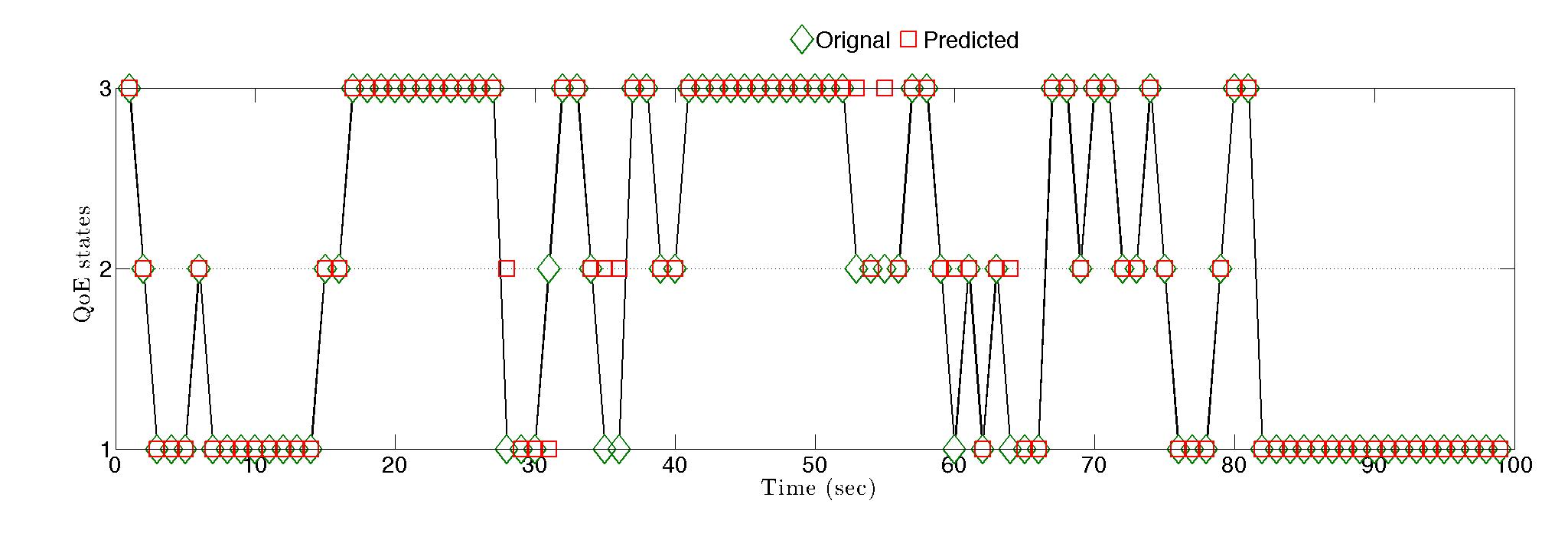}\caption{Figure showing HMM and M-MIP based method can accurately predict QoE
in case of WLAN congestion.}
\end{figure}

\begin{figure} \begin{centering} \includegraphics[scale=0.25]{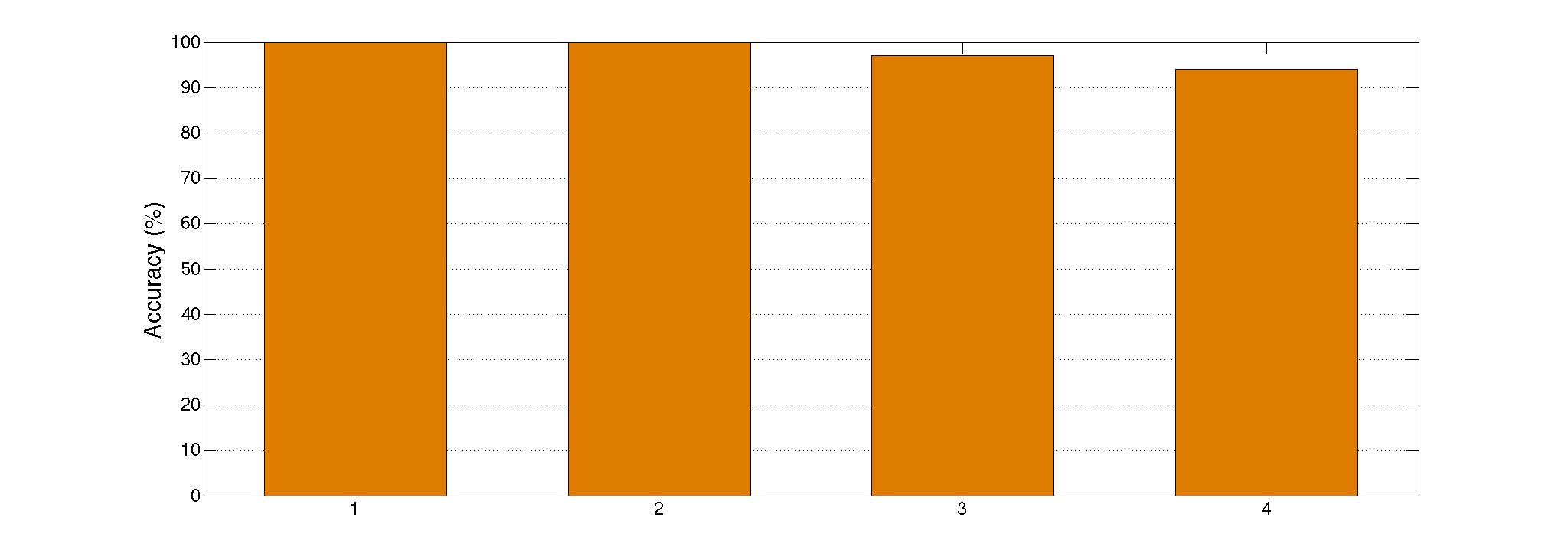}\caption{Prediction accuracy using HMMs for WLAN and CDMA2000 network interfaces using experimental data set. 1.) WLAN with ITU-T G.711 codec. 2.) WLAN with ITU-T G.729 codec. 3.) CDMA2000 with ITU-T G.711 codec. 4.) CDMA2000 with ITU-T G.729 codec.} \par\end{centering} \end{figure}

\begin{figure}
\centering{}\includegraphics[scale=0.22]{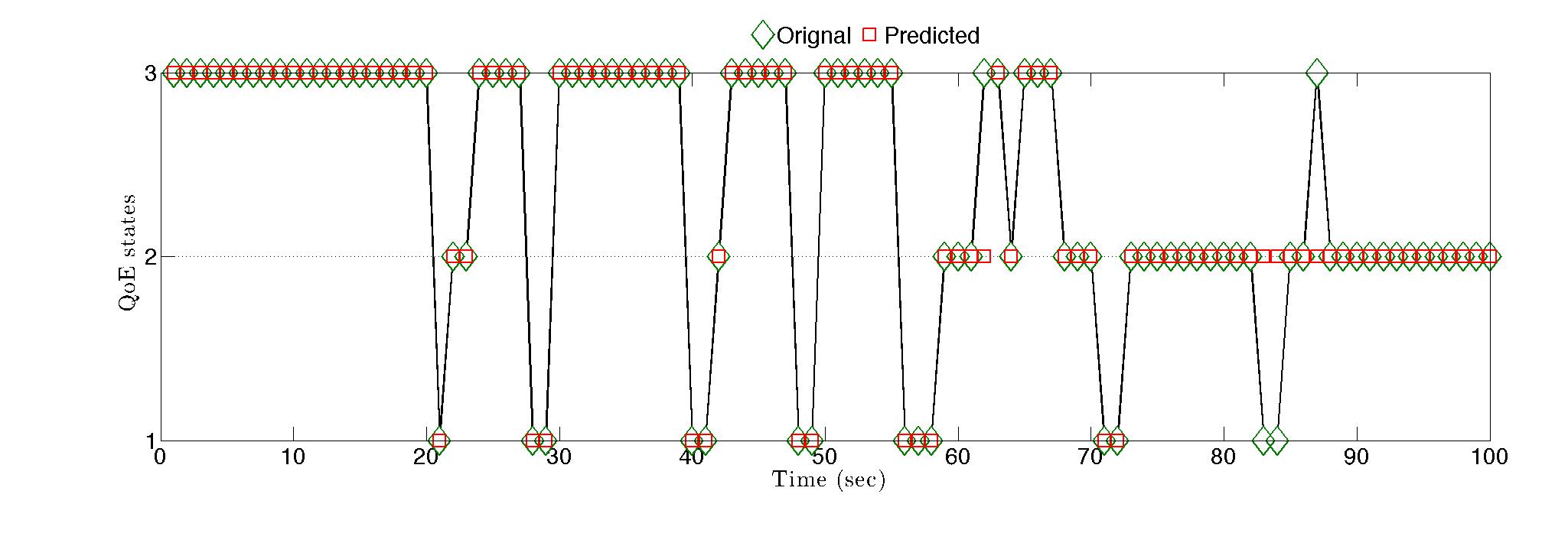}\caption{Figure showing HMM and M-MIP based method can accurately predict QoE
in case of cellular network.}
\end{figure}

\subsection{Proactive QoE Provisioning using Reinforcement Learning in Heterogeneous
Access Networks}

The previous section concluded that HMMs can accurately predict users'
QoE in HANs. We now extend our HMM based approach to facilitate proactive
QoE-aware handoffs in HANs. In particular, our agent, the MN, predicts
the QoE state using HMMs and then selects an optimal action (regarding
handoffs) based on the Q-value function as shown in Algorithm 1.

\subsubsection{Minimizing the Number of Handoffs Using HMMs and Reinforcement learning}

To validate our proposed approach, we compare our results with Multimedia
Mobility Manager ($M^{4}$) \cite{m4} and a naive scheme similar
to multi-attribute decision making (MADM) methods \cite{Balasubramaniam2004,nasser2006,wang2012}.
We chose the \emph{number of handoffs} as a criterion for results
analysis. $M^{4}$ considers the relative network load (RNL) metric
\cite{tlm} for network load-aware handoffs. The RNL metric considers
the running average of RTT jitter values to estimate the load on access
networks. The RTT values are computed based on the BU/BA pair sent
between the MN and HA/CN. The RNL metric is computed as follows: 

\begin{equation}
RNL=Z_{n}+cJ_{n}
\end{equation}

\begin{equation}
Z_{n}=\frac{1}{h}RTT_{n}+\frac{h-1}{h}Z_{n-1}
\end{equation}

\begin{equation}
RTT_{n}=R_{n}-S_{n}
\end{equation}
\begin{equation}
D_{n}=RTT_{n}-RTT_{n-1}
\end{equation}

\begin{equation}
J_{n}=\frac{1}{h}|D_{n}|+\frac{h-1}{h}J_{n-1}
\end{equation}
 where, $S_{j}$ is the time to send the BU packet $j\in n$ from
the MN to HA/CN and $R_{j}$ is the time of arrival of the BA packet
(corresponding to packet $j)$ from the HA/CN to the MN. $h$ is the
history window for calculating the weighted average, where $h=5$
is considered to be an optimal value \cite{m4,SLMTELSYS2008}. $c$
represents the weight of the RTT value compared to the RTT jitter
value. For example, if $c=5$, it means that the RTT jitter value
contributes 5 times more than the RTT value. Finally, the variables
$Z,$ $D$ and $J$ are initialized as: $Z_{0}=RTT_{0}$; $D_{0}=0$;
and $J_{0}=D_{1}.$ The network with lower RNL value will be the target
for handoff. $M^{4}$ using the RNL metric assumes that handoffs based
on network load prediction will be beneficial for a wide range of
applications, especially real-time applications such as VoIP. 

As mentioned previously, a naive scheme on the other hand, is similar
to MADM methods \cite{Balasubramaniam2004,nasser2006} and aims at
selecting a network providing best QoS at a particular time $t$.
Thus, for example, a network $i$ providing best QoS in terms of bandwidth
$(B_{i})$, delay $(D_{i})$, jitter $(JIT_{i})$ and packet loss
ratio $(PLR_{i})$ will be the target for handoff. The QoS function
for a particular network $i$ can be written as \cite{nasser2006}:
\begin{equation}
QoS_{i}=w_{B}*B_{i}+w_{D}*\frac{1}{D_{i}}+w_{JIT}*\frac{1}{JIT_{i}}+w_{PLR}*\frac{1}{PLR_{i}}
\end{equation}
where $i\in\, I$$;$ $w_{B},w_{D},w_{JIT}\, and\, w_{PLR}$ are the
set of weights; and $w_{B}+w_{D}+w_{JIT}+w_{PLR}=1$. The network
$i$ with higher $QoS$ value will be the target for handoff. 

To compare our method with $M^{4}$ and the naive method, we formulated
a hypothesis that our method should reduce the expected number of
vertical handoffs while maximizing user's QoE. Our hypothesis was
based on the fact that $M^{4}$ and the naive methods considers precise
RNL and QoS values for handoffs. In both methods, even the very close
RNL or QoS values on network interfaces can cause sudden handoffs.
\emph{Using these methods, handoffs can occur even if the QoE level
is same on both network interfaces, which might not be optimal}. On
the contrary, our method considers finite states to represent QoE
where the HMMs directly learns and predicts QoE. Further, the Q-value
function learns the optimal action-value pair for proactive QoE-aware
handoffs. We now consider the problem of reducing expected number
of handoffs $(\Delta)$ between two network interfaces (WLAN and CDMA
2000) while maximizing user's QoE. For a vertical handoff decision
problem, we used the reward function $(\Re)$ defined in Eq. 5. To
avoid ping-pong effects during handoffs, we chose the same hysteresis
margin proposed in \cite{m4} and considered ITU-T G.711 and ITU-T
G.729 voice codecs for results analysis. In the Q-learning algorithm
(Eq. 15), we selected $\alpha=0.80$ and $\gamma=0.95$. 

For results analysis, we again considered the experimental data set
from the authors in \cite{m4} (see section 6.2.1 for the discussion
on experimental setup). In this paper, our proposed method considers
delay values to predict QoE. Further, $M^{4}$ also considers the
delay values to compute the RNL metric. Thus, for comparative analysis,
the naive method only considers delay $(D_{i})$ determine $QoS_{i}$.
The table 5 shows that the proposed method almost matches the best
case method where our method reduces the average number of unnecessary
handoffs by almost 56\% compared to $M^{4}$. Similarly, compared
to the naive scheme, the proposed method reduces the average number
of handoffs by approximately 65\%. The difference in the number of
handoffs for ITU-T G.711 and ITU-T G.729 codec is due to the fact
that ITU-T G.729 codec provided better QoE and could sustain variations
in RTTs better than the ITU-T G.711 codec. The proposed method in
case of the ITU-T G.711 codec tried to maintain higher QoE levels
by making higher number of handoffs to the network interface that
provided higher QoE. The results presented in this section validates
that: 
\begin{enumerate}
\item Our passive probing method based on HMMs can be used by any MN incorporating
multi-homed mobility management protocol such as M-MIP for efficient
QoE prediction for all network interfaces simultaneously. Our method
eliminates the need for active probing mechanism. Instead, it uses
signaling mechanisms such as BU/BA packets for accurate QoE prediction.
We showed that the average prediction accuracy of approximately 97\%
was achieved.
\item Our method incorporating HMMs and RL can be used by any MN for seamless
roaming in HANs. Our results clearly validate that our method reduce
the average number of handoffs by approximately 65\% compared to the
state-of-the-art methods.
\begin{table}
\caption{Results showing expected number of handoffs reduced by proposed scheme
compared to Naive scheme and $M^{4}$ .}

\centering{}%
\begin{tabular}{|r|c|c|c|c|>{\centering}p{0.1\columnwidth}|>{\centering}p{0.1\columnwidth}|}
\hline 
Cases & Best & Naive & $M^{4}$ & Proposed & Reduction against Naive  & Reduction against $M^{4}$\tabularnewline
\hline 
\hline 
ITU-T G.729 & 21 & 63 & 52 & 21 & 66.67\% & 59.61\%\tabularnewline
\hline 
ITU-T G.711 & 21 & 64 & 52 & 24 & 62.50\% & 53.84\%\tabularnewline
\hline 
\end{tabular}
\end{table}
 
\end{enumerate}
\subsection{Discussion}
Developing metrics for users' QoE prediction is an active area of research. Recently, several methods for QoE measurement and prediction were proposed \cite{barakovic2013survey,mitraqoereview14}. For example, the ITU-T E-Model \cite{G.107}, PESQ \cite{pesq}, PSQA \cite{psqa}, Menkovski \textit{et. al} \cite{menkovski2009optimized}, and CaQoEM \cite{mitratmc2014}. In this paper, we considered QoE as the function of the ITU-T E-Model \cite{G.107} as it is the most widely used metric for users' QoE prediction in both industry and academia.  The E-Model outputs the mean opinion score as a non-linear function of delay, packet loss and numerous other parameters including codec-type, mean loss burst length and advantage factor. This parameter however, does not consider user QoE ratings over time and is limited to VoIP applications. We believe, capturing users' QoE over time can help build robust and more realistic models. As part of the future work, we would like to incorporate SCaQoEM \cite{mitradbn} for users' QoE estimation. We can then try to integrate PRONET with SCaQoEM for highly adaptive handoffs for various applications used under varying context. 

We would also like to integrate mechanisms for users ratings collection on-the-fly, once the MN goes through the handoff. This way the MN can optimize the reward function based correct/incorrect actions taken by it based on direct users' feedback. Finally, in this paper, we considered two network interfaces (WLAN and CDMA2000). In the future, a MN may incorporate several other network interfaces such as Bluetooth and ZigBee. Thus, as mentioned in section 3, for each network interface, a new HMM will have to be learnt.  Similarly, for Q-learning, new states and actions for the new network interface will have to be incorporated. However, this can easily be integrated within our system.

As mentioned previously, QoE prediction and provisioning is a challenging task. Till date, nearly all methods \cite{piamrat2008,piamrat2011,varela2011,Rosario2013,Quadros2013} treated these problems independently \cite{varela2011,mitraatnac}. Therefore, lacking either of the two crucial functionalities without which an efficient system for QoE provisioning cannot be realized.  To the best of our knowledge, ours is the first paper to integrate QoE prediction and proactive QoE provisioning capabilities into a mobility management protocol. In all, the results presented in the previous subsection clearly validate our methods for efficient QoE prediction and proactive QoE provisioning in HANs. We showed that the HMMs trained using passive probing (thereby, eliminating the need for additional probe packets) i.e., using the BU/BA messages exchanged using between the MN and the HA can be used for accurate QoE prediction with high accuracy. Further, compared to the state-of-the-art methods, using Q-learning, a large number of handoffs can be reduced significantly.

\section{Conclusion}

This paper proposed, developed and validated a novel method for QoE
 prediction using HMMs and M-MIP based passive probing mechanism.
Our results based on simulations and experimental studies conclude
that HMMs trained using OWD or RTT delay calculated using BU and BA
packets are suitable to accurately estimate and predict QoE for VoIP
applications. Our method achieves an average prediction accuracy of
97\%.
The highlight of our approach is that it learns QoE states automatically
and quickly by considering finite state spaces compared to the state-of-the
art. 

This paper also proposed, developed and validated a novel reinforcement
learning based method for proactive QoE-aware handoffs in HANs. We
demonstrated that our method reduces a number of vertical handoffs
by 60.65 \% compared to $M^{4}$ and a naive method while maintaining
desired QoE. As an immediate future work, we will incorporate our methods on
mobile devices running Android operating system and on other resource constraint devices such as Raspberry Pi.

\bibliographystyle{ieeetr}
\bibliography{JMD_PHD_REFERENCES,JMD_BHARAT_PHD_REF}

\end{document}